\documentclass[12pt,preprint]{aastex}
\usepackage{multirow}

\begin{document}

\slugcomment{\it Accepted for publication in ApJS on 1 September 2007.}

\title{CGRaBS: An All-Sky Survey of Gamma-Ray Blazar Candidates}

\author{Stephen E.\ Healey\altaffilmark{1}, Roger W.\ Romani\altaffilmark{1},
Garret Cotter\altaffilmark{2}, Peter F.\ Michelson\altaffilmark{1}, \\
Edward F.\ Schlafly\altaffilmark{1}, Anthony C.\ S.\ Readhead\altaffilmark{3},
Paolo Giommi\altaffilmark{4}, \\
Sylvain Chaty\altaffilmark{5}, Isabelle A. Grenier\altaffilmark{5},
Lawrence C. Weintraub\altaffilmark{3}}

\altaffiltext{1}{Department of Physics/KIPAC, Stanford University, Stanford, CA 94305; sehealey@astro.stanford.edu, rwr@astro.stanford.edu.}
\altaffiltext{2}{Department of Astrophysics, University of Oxford, Oxford OX1 3RH, UK}
\altaffiltext{3}{Department of Astronomy, California Institute of Technology, Pasadena, CA 91125}
\altaffiltext{4}{ASI Science Data Center, I-00044 Frascati, Italy}
\altaffiltext{5}{Laboratoire AIM, CEA/DSM - CNRS - Universit\'{e} Paris Diderot, 
Service d'Astrophysique, CEA Saclay, 91191 Gif/Yvette, France}

\begin{abstract}
	We describe a uniform all-sky survey of bright blazars, selected primarily by their
flat radio spectra, that is designed to provide a large catalog of likely
$\gamma$-ray AGN.  The defined sample has 1625 targets with radio and X-ray properties similar to 
those of the EGRET blazars, spread uniformly across the $|b|>10^\circ$ sky.  We also report
progress toward optical characterization of the sample; of objects with known $R<23$, 85\% have been
classified and 81\% have measured redshifts.  One goal of this
program is to focus attention on the most interesting (e.g., high redshift, high luminosity, ...)\
sources for intensive multiwavelength study during the observations by the Large Area Telescope
(LAT) on {\it GLAST}.
\end{abstract}


\section{Introduction}

It is well-known \citep{3eg,mattox} that many of the high-latitude EGRET sources 
are associated with the bright, flat radio spectrum AGN known as blazars. 
Sowards-Emmerd et al.\ (2003; SRM03) quantified such associations, developing a 
combined figure of merit (FoM), which measured the likelihood that an individual
radio/X-ray source near the large ($\sim$$0.7^\circ$) Third EGRET Catalog 
(3EG, \citealt{3eg}) uncertainty regions is the $\gamma$-ray counterpart.
They also noted that there are many radio-loud blazars with very
similar properties not obviously associated with a 3EG source. A likely explanation
is that blazars are very variable at high energy, with duty cycles for the bright, flaring
state as small as a few percent \citep{hart93,kniffen}. During the limited 
(typically two weeks per pointing direction) 3EG exposure many of these sources may have
been in quiescence. Accordingly, \citet{srm05} extended the SRM03 analysis 
by selecting ``3EG-like'' blazars, i.e., sources whose radio flux density and spectrum 
(and X-ray flux) were very like those of the 3EG blazars but which happened not to 
lie within a 3EG test statistic (TS) uncertainty region. The positions of these
sources showed a clear excess of $\gamma$-ray photons over background and these sources
are likely to show $\gamma$-ray high states during future missions.

	The Large Area Telescope (LAT) on {\it GLAST} will provide an improvement of several
orders of magnitude over EGRET/{\it CGRO} with an increased sensitivity in the 50 MeV $-$ 300 GeV
range and a wide ($>$2.5 sr) field of view. The LAT should detect
many thousands of sources during the 5- to 10-year mission, with a large fraction of the
high-latitude sources being blazars.  The early mission will be devoted to a sky survey, covering
the entire sky at good sensitivity every three hours. This will greatly
enhance the likelihood of detecting transient and variable sources, such as
blazars.  While several large samples of blazars have been compiled
recently (see especially the ASDC blazar catalog, \citealt{asdc}, and the ROXA catalog,
\citealt{roxa}), there is a
surprisingly incomplete knowledge of the radio-bright, flat-spectrum population,
which is most clearly associated with the GeV $\gamma$-ray sources.  We seek
to rectify this by defining CGRaBS, the \underline{C}andidate \underline{G}amma-\underline{Ra}y
\underline{B}lazar \underline{S}urvey, a large sample of EGRET-like blazars selected across
the extragalactic sky.  By obtaining optical classifications and redshifts for a large
fraction of these sources, we plan to enable prompt, intensive follow-up of the most
interesting (e.g., high redshift, high luminosity, peculiar spectrum) sources that {\it are}
detected in the LAT sky survey data. Furthermore, identification of a
substantial fraction of the LAT sources with blazars will allow us
to focus on the non-blazar remainder, potentially isolating new classes of
cosmic $\gamma$-ray emitters.

\section{Sample Selection}

	For any FoM-type counterpart selection, it is important to have uniform
parent populations.  \citet{crates} have recently developed such a catalog, CRATES,
which extended results of the CLASS survey \citep{class} to obtain
8.4 GHz observations of all $|b|>10^\circ$ objects brighter than 65 mJy at 4.8 GHz
with spectral indices $\alpha > -0.5$ (where $S \propto \nu^\alpha$). To estimate
the radio spectral index of the core, we use the NVSS \citep{nvss} and
SUMSS \citep{sumss} lower-frequency surveys.  The result is
a sample of over 11,000 flat-spectrum radio sources with interferometric measurements
at $\sim$1 GHz and 8.4 GHz (with FWHM beam sizes $\sim$$40\arcsec$ and $\sim$$0.25\arcsec$
respectively), giving precise positions, spectral indices, and morphologies
for the compact components.  The CRATES catalog is as uniform as possible for the high-latitude
($|b|>10^\circ$) sky, limited by gaps in which the initial 4.8 GHz data are unavailable.
We believe that this catalog is an excellent starting point for
comparison with other all-sky samples (e.g., microwave and $\gamma$-ray).

	Here we wish to find EGRET-like blazars, so we adopt the
FoM of SRM03, which was derived from comparing the well-established 3EG 
blazar sources with the northern (CLASS-generated) subset of
the CRATES catalog. This FoM is given by the heuristic fitting formula 
$\mathrm{FoM_{3EG}}=100 \times P_\alpha \times P_S \times P_{\mathrm X} \times P_{\mathrm{TS}}$, where the $P$
terms are ``excess probabilities'' for the observed parameters for radio sources near 
3EG sources.  Here, $P_\alpha = 0.19-0.35\alpha_{\mathrm{low}/8.4}$ ($0 \le P_\alpha \le 0.4$),
$P_S=-3.47+2.45 \log_{10} S_{8.4} - 0.34 \log_{10}^2S_{8.4}$ ($0 \le P_S \le 1$)
and $P_\mathrm{X} = 0.99 + 0.41 \log_{10} F$ ($0.5 \le P_\mathrm{X} \le 1$), with $F$ the RASS \citep{rass}
counts per second and the $P$ terms bounded to the ranges in parentheses. Finally,
$P_{\mathrm{TS}}=1-\mathrm{CL}$, where CL is the confidence limit of the 3EG source
localization contour passing through the position of the radio source.
In essence, the FoM is
composed of the product of the ``excess'' probabilities of sources of a given flux density, spectral
index, etc.\ over random chance. While the FoM probability is not directly normalized,
``false positive'' rates were computed at each FoM level by comparison with the statistics of
scrambled versions of the sky catalogs.
Of course, once we have an initial survey of LAT blazar sources, it will be 
appropriate to derive new coefficients, ``re-training'' the FoM against this sample. 

        To develop an all-sky survey of blazar candidates, we compute an FoM for
each source in the CRATES catalog. We must do this without reference
to 3EG sources. Thus for this paper we define $\mathrm{FoM}=  P_\alpha \times P_S \times P_\mathrm{X}$.
To connect with $\mathrm{FoM_{3EG}}$, note that a blazar with the present $\mathrm{FoM}=0.2$ would
correspond to a $\mathrm{FoM_{3EG}}=1$ at the 95\% localization contour of a $\gamma$-ray
source, a ``likely'' ($>$90\% correct) identification.
With this definition, 5059 of the CRATES sources have a nonzero FoM.
To focus our follow-up on the best and most 3EG-like objects, we define CGRaBS as
those 1625 sources with $\mathrm{FoM} > 0.04$. This corresponds to an SRM03 value of
$\mathrm{FoM_{3EG}}=2$ at the $50\%$ localization contour, a very likely association,
and a $\mathrm{FoM_{3EG}}=0.2$ for a source at the 95\% confidence contour, a reasonable 
($>$80\%) likelihood of an association.
Figure 1 shows an Aitoff equal-area projection of the CGRaBS sample along with its parent
survey, CRATES.  Figure 2 shows a projection of the CGRaBS sample indicating the FoM of
each source.

\begin{figure}[h]
\begin{center}

\includegraphics[width=\textwidth,angle=0,keepaspectratio=true]{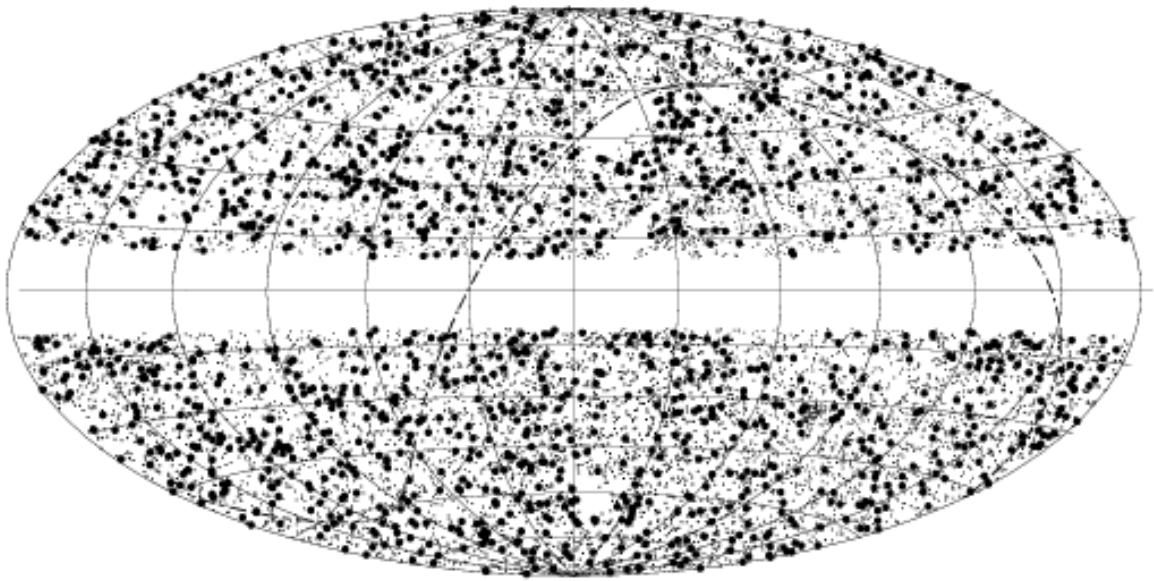}

\caption{Aitoff equal-area projection of the CRATES parent sample (small dots) and the
CGRaBS sample (large dots) in Galactic coordinates $(l,\,b)$.  The central meridian is
$l = 0^\circ$. A few small holes are visible just below $\delta=0^\circ$ (dot-dashed line),
stemming from incomplete PMN sky coverage.}
\label{aitoff1}
\end{center}
\end{figure}

\begin{figure}[h]
\begin{center}

\includegraphics[width=0.8\textwidth,angle=0,keepaspectratio=true,trim=1.4in 0.8in 1in 3.75in]{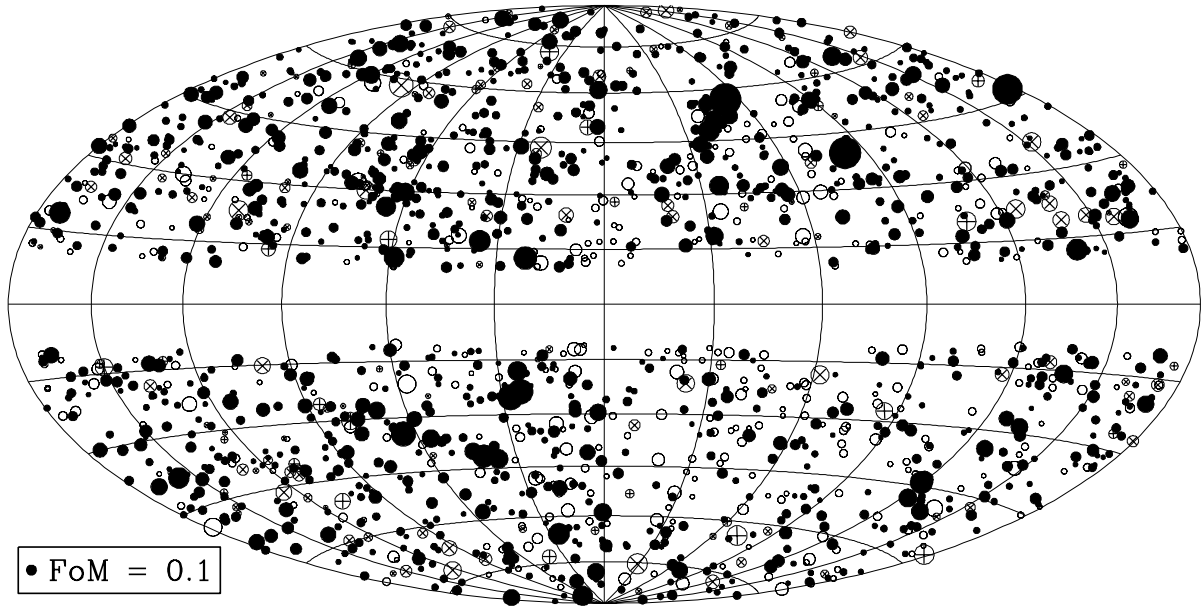}

\caption{Aitoff equal-area projection of the CGRaBS sample in Galactic coordinates $(l,\,b)$.
The central meridian is $l = 0^\circ$.  The radius of each dot is proportional to the FoM of the
source; the dot for a source with FoM = 0.1 is shown for comparison.  The dot styles indicate
optical classifications (see \S\S3.2.1-2): $\bullet$ = FSRQ, $\otimes$ = BLL, $\oplus$ = AGN,
$\circ$ = unknown.}
\label{aitoff2}
\end{center}
\end{figure}

	The radio spectral index is a major component of our FoM; thus, since the
interferometric observations at 8.4 GHz and low frequency were non-simultaneous, variability
can, in principle, affect our FoM measurements. Luckily, the variability in the radio
is modest compared to the high energy bands: \citeauthor{crates} found that the mean
8.4 GHz variability is $\le$14\% and the low-frequency variability on the relevant
several-year timescale is even smaller. Thus, we do not expect that radio variability
will dramatically affect our FoM estimates. Further, the (more likely variable) RASS
detections turn out not to be a major selection bias in this survey.
If the X-ray contribution to the FoM is ignored and a purely radio-based FoM
is computed, then 98.5\% of our sources still satisfy the CGRaBS FoM cutoff.  Thus, while
the X-ray flux from a small number of sources boosts them into the sample, the main effect
of the X-ray contribution is to shuffle the ranking within the set of sources that are
already qualified. Since the radio FoM weighting increases for bright and inverted (rising)
spectra, its net effect is to impose an effective {\it extrapolated} flux density limit at
a higher radio frequency.  For example, the FoM = 0.04 cutoff corresponds to an
extrapolated flux density at 100 GHz of $S_{100} > 230$ mJy (although we do not
expect all sources to have a constant $\alpha$ to such high frequency). Less than 
1\% of the full CGRaBS targets have an extrapolated flux below this threshold, 
and these are all low-FoM sources with very high 
X-ray flux (i.e., largely high-peaked sources; see the next section).

	Three CGRaBS sources warrant special comment.  The CRATES entry for J0352$-$2514 is
a combination of 8.4 GHz observations at two epochs, one with an unflagged mapping error and a grossly
erroneous position.  The CGRaBS entry for J0352$-$2514 uses only the good epoch to determine the correct
position, the 8.4 GHz flux density, and the spectral index.  Sources J0805$-$0111 and J1639$+$1632 have
nominal CGRaBS spectral indices (and thus FoMs) that are almost certainly overestimates.  Their NVSS
counterparts have marginally resolved jet structure, and the NVSS decompositions offset the core toward
the jet.  A faint, spurious counter-jet component was introduced and, being slightly closer to the 8.4
GHz position, was selected as the 1.4 GHz counterpart, leading to a highly inverted spectral index and a high FoM.
We include these sources in the survey since they satisfy the CGRaBS prescription; a more careful treatment of
the NVSS counterparts would give a smaller spectral index and FoM.  This effect is quite rare, occurring in CGRaBS
for only these 2 sources (out of 1625, or 0.12\%) and in CRATES for no more than 20 sources (out of 11,131,
or 0.18\%).

\section{Optical follow-up}

	We have specifically {\it not} required a previous optical (or X-ray) detection of
our blazar candidates. This radio-driven selection allows us to sample completely the
flat-spectrum sources and avoid biasing the detected population.
For example, X-ray--bright sources are preferentially low-power ``blue'' blazars
such as BL Lacs (so-called high-peaked blazars, or HBLs; \citealt{hbl}).  Similarly, requiring
optically bright counterparts can bias
the sample toward low redshift.  However, since the principal goal of the CGRaBS project is
to secure optical identifications, we do need good magnitude estimates. To
maximize uniformity, we are working toward complete identification for $R < 23$.
In practice, we have also observed a number of radio-bright and
X-ray/$\gamma$-ray--bright but optically faint sources beyond this limit to explore the
extrema of the population.

\subsection{Counterparts and photometry}

One defining blazar characteristic is rapid optical variability. Thus, we must set a fiducial ``epoch''
for the optical magnitudes. In practice, we use the USNO-B1 catalog \citep{usno}
since this is the largest source of suitable $R$ magnitudes; we take the fiducial magnitude
to be that of the more sensitive second epoch survey (R2).  Since we have precise radio
positions for the cores of all sources, we identify a USNO-B1 source as the
counterpart of a CGRaBS source if the optical position is within $1.5\arcsec$ of the radio
position. This gives a large fraction of the required magnitudes, with completeness dropping
between $R\sim 20$ and $R\sim21$.
For the north Galactic cap, we can supplement these with SDSS identifications
(through Data Release 5, \citealt{sdss}) to $r^\prime \la 22$. In
confused cases, these archival data were examined visually to determine the best counterpart
match.  In a number of cases, we were also able to see clear counterparts that were
too faint for inclusion in the USNO-B1 catalog but whose magnitudes could be reasonably estimated
by measurement of the digitized plate data.  In view of the variable blazar
magnitudes and non-stellar colors, this low-precision photometry is adequate
to guide the follow-up spectroscopy.

	To complete the process of optical identification (and to improve a few poor USNO-B1
magnitudes), we have conducted our own imaging campaign, primarily at the 5 m Hale 
Telescope at Palomar, the 3.6 m New Technology Telescope (NTT) at La Silla, and the 2.7 m
Harlan J.\ Smith Telescope at McDonald.  Typical
exposures were 180 s through Gunn $r^\prime$ under varying conditions, and magnitudes were
calibrated against multiple field stars.  For some particularly
interesting sources (e.g., high radio-to-optical flux ratio candidates for high redshift),
these were supplemented with {\it izJHK} imaging.  We do not report here on
these optical/IR SEDs.  All $r^\prime$ magnitudes have been converted to $R$
using the average color term ($R-r^\prime=- 0.253$) of CGRaBS sources detected by
both the SDSS and USNO-B1.  A magnitude (or limit) for each source is listed in Table 2. 
For some of the lowest redshift sources, the magnitude is dominated by the flux from 
the (extended) host galaxy.  We also list the nominal Galactic extinction for
the source direction $A_R$, derived from the \citet{schleg} maps. Even though the sources are at
high latitude, there are a few targets behind dust clouds, indicating a large nominal
extinction. However, we do not expect extinction to bias our measured population as
the large $A_R$ are not preferentially associated with the faint targets. Furthermore,
only 4\% of the blazars have $A_R>1$ and 0.5\% have $A_R>2$; only 4 sources are
excluded from the targeted $R=23$ sample by the known extinction.
As of 2007.5, there are 88 objects
(5.4\%) that do not have measured $R$ magnitudes; of these, 45 have limits fainter than
$R=23$ and thus do not nominally require spectroscopy for the complete survey.  The sources
with brighter limits will be the subject of further imaging.  Note that with 68 CGRaBS
sources known to be fainter than $R=23$, we expect that the survey will be $>$95\% complete at this
magnitude limit.

	Figure 3 shows the distribution of $R$ magnitudes and limits. 
Since the $A_R$ are in general small,
the extinction-corrected histogram is very similar.  At first sight,
the rapid drop between $20<R<21$ would seem to be due to the USNO-B1 survey completeness limit.
However, we have sufficient deeper CCD imaging to determine that the drop in numbers is largely
intrinsic, although we need to complete the imaging before we can characterize the details
of the faint source distribution. The right panel shows that we need to complete
identifications to faint magnitudes ($R>19$) to get a representative sample of the higher redshift
sources.

\begin{figure}[ht]
\centering
\plottwo{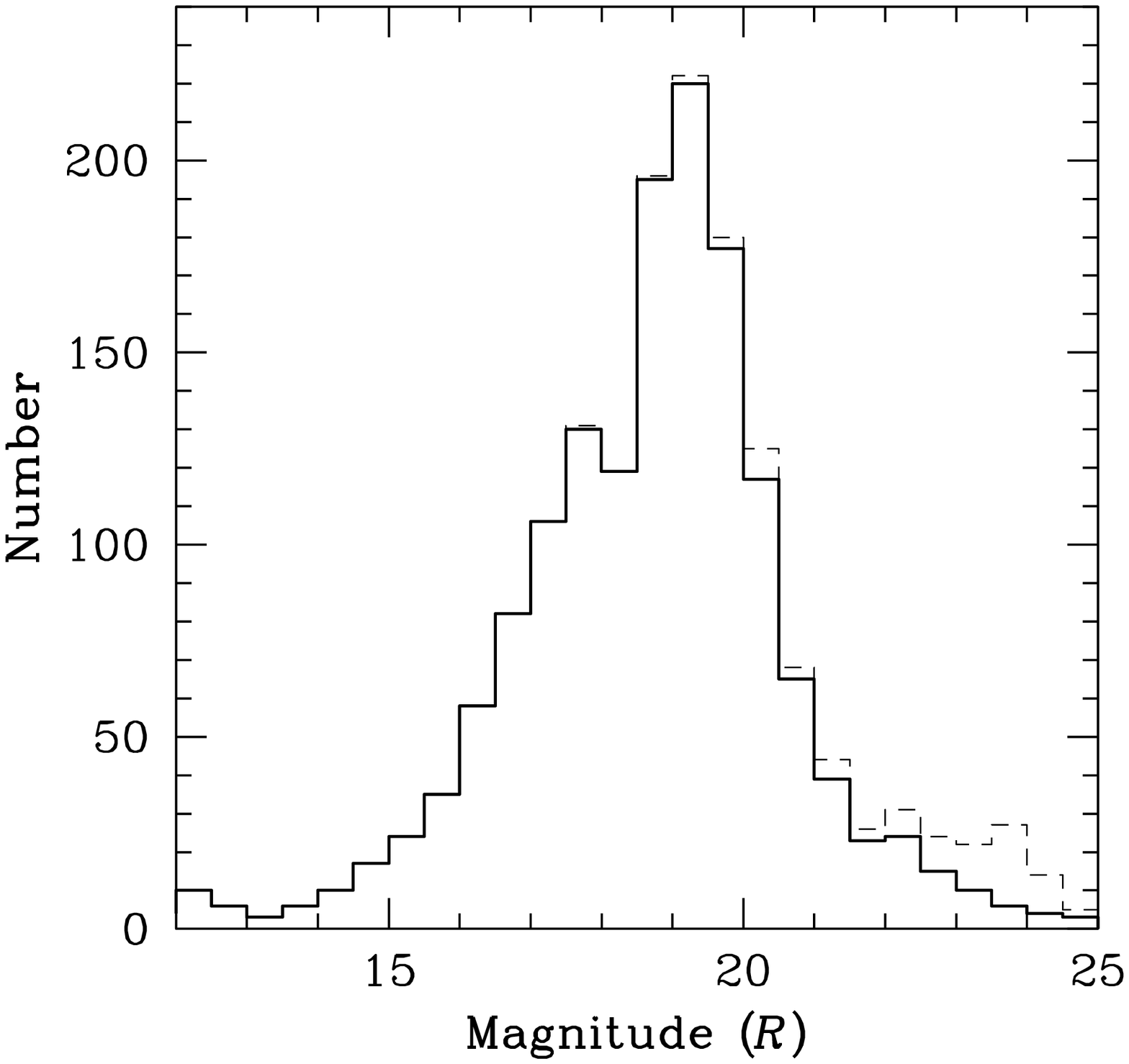}{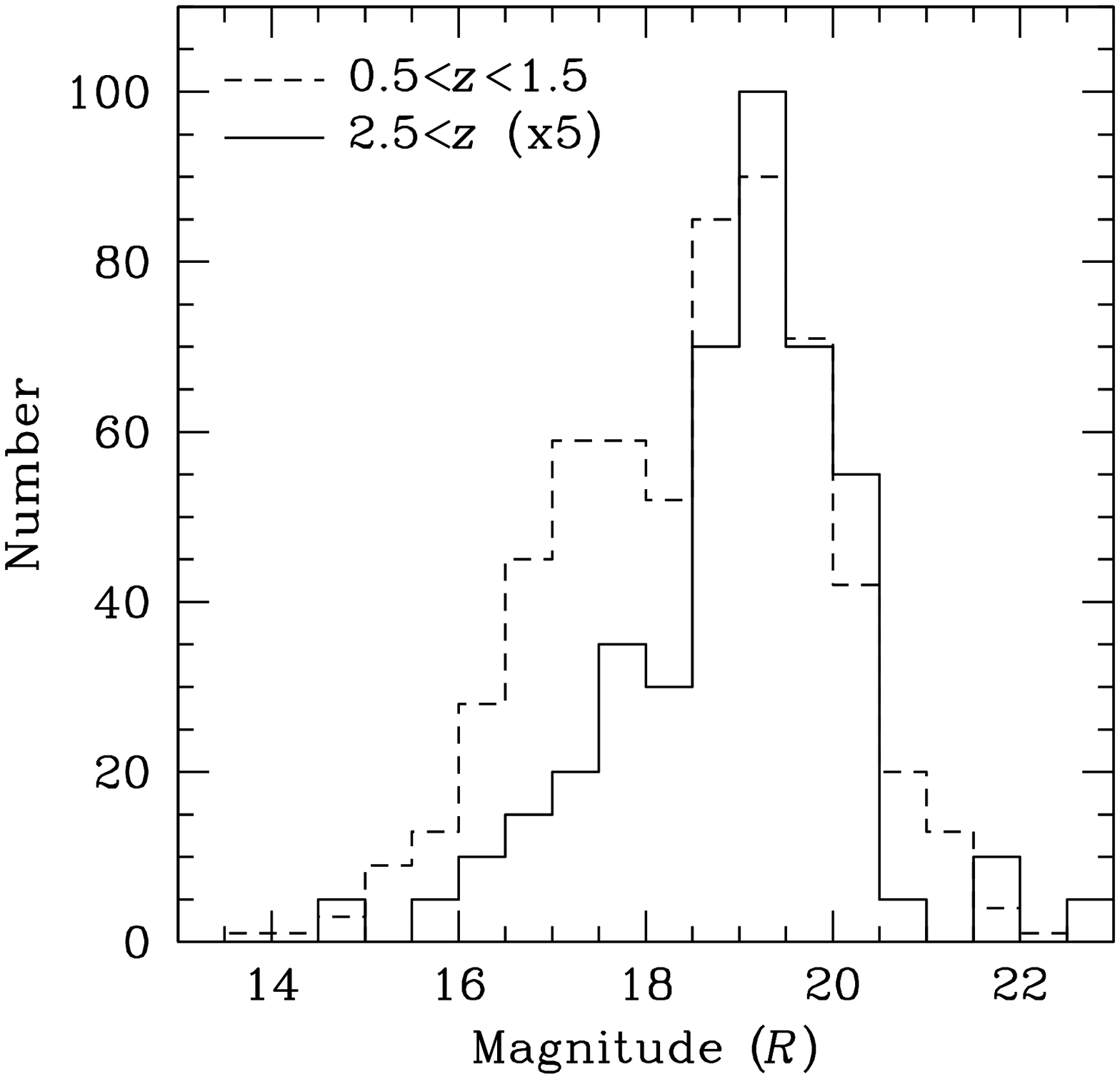}
\label{magplots}
\caption[magplots]{{\it Left:} Magnitude distribution of CGRaBS sources.  Lower limits on $R$ are
shown by the dashed histogram.  {\it Right:} Magnitude distributions for low-redshift and
high-redshift sources.}
\end{figure}

\subsection{Spectroscopy}

	Our spectroscopic goals are a basic classification of the AGN type, redshift measurement,
and measurement of emission line equivalent strengths and kinematic widths (for luminosity 
and mass evolution studies).  Thus, the bulk of our new observations have been low-resolution
$\mathcal{R} \sim 500$$-$1500 long-slit spectroscopy.  Most of the sources are flat-spectrum
radio quasars whose broad lines allow easy identification with relatively low signal-to-noise.
However, a significant fraction of the sources ($\sim$15\%) are weak-lined BL Lac sources. For
these, we require high S/N and/or high resolution to determine the redshift from host
absorption lines.  Such measurements require long exposures with large telescopes. At present,
we have identified sources as BL Lacs to $R\sim 20$, but our ability to measure the redshift
drops significantly above $R\sim 18.5$; these BL Lacs are the subject of further spectroscopy
at higher dispersion.  In this paper, we present a progress report on the optical identifications.
Additional papers will discuss the properties of the complete sample, the source SEDs,
and the constraints on blazar evolution.

\subsubsection{Observations}

	A fair fraction of the CGRaBS sources are bright, well-known AGN; thus, we
have vetted our catalog against the twelfth edition of the V\'{e}ron quasar catalog \citep{veron}
and the SDSS DR5 quasar catalog \citep{sdssq}.  We have also queried
NED ({\tt http://nedwww.ipac.caltech.edu/}) for all CGRaBS sources to find any other redshifts
and identifications in the literature.
Archival data identify $\sim$45\% of the CGRaBS objects ($\sim$60\% of the
redshifts in hand); the remainder are the targets of our own spectroscopic campaigns.
The great workhorse of our spectroscopic effort has been the 9.2 m Hobby*Eberly Telescope (HET)
at McDonald, which has observed hundreds of CGRaBS sources in the accessible declination
band $-11^\circ < \delta < +73^\circ$.  The telescope is fully queue-scheduled \citep{shetrone},
allowing  us to receive data remotely year-round and to spread the cost of inclement weather and 
unfavorable conditions.  We use the Marcario Low Resolution Spectrograph (LRS; \citealt{lrs}) 
with grism G1 (300 lines mm$^{-1}$), $2\arcsec$ slit, and a Schott GG385 long-pass filter
for a resolution of $\mathcal{R} \approx 500$.  Typical exposures are $2\times 600$ s,
providing redshifts of emission-line objects to $R \sim 22$; brighter objects are 
also observed under poor conditions with $2\times 300$ s.

	In addition to our ongoing HET observations, we have mounted dedicated campaigns at 
a number of other facilities.  We conducted three runs totaling 13 nights (over half lost
to weather) on the 2.7 m Harlan J. Smith Telescope at McDonald, using 
the Imaging Grism Instrument (IGI) and the 6000 \AA\ VPH grism.  We observed 28
objects with the 1.5 m telescope at Cerro Tololo in the ``13/I'' setup (grism 13, 150 lines
mm$^{-1}$) in service observing mode as part of the Small and Medium Aperture
Research Telescope System (SMARTS) program.  We conducted two runs totaling 8 nights on the
3.6 m NTT at La Silla with the ESO Multi-Mode Instrument (EMMI) in the low-resolution
spectroscopy (RILD) mode and grism 2 (300 lines mm$^{-1}$).  To date, we have had three runs
totaling 12 nights on the 5 m Hale Telescope at Palomar with the double
spectrograph (DBSP), using a 300 lines mm$^{-1}$ grating on the blue side and a 316 lines
mm$^{-1}$ grating on the red side.  We observed 12 objects with the 8.2 m Kueyen telescope
(the second unit telescope at the Very Large Telescope, VLT) in service observing mode with Focal
Reducer/Low-Dispersion Spectrograph 1 (FORS1) and grism GRIS\_300V (300 lines mm$^{-1}$).  Finally,
we have had three runs totaling 4 nights on the 10 m Keck I Telescope at Mauna Kea (however,
the night of 2006 October 28 was the first observing night after the 2006 earthquake, and pointing
was severely restricted; observations remained substantially constrained even on the night of
2006 November 24).  For these observations, we used the Low Resolution Imaging Spectrometer (LRIS),
employing a 600 lines mm$^{-1}$ grism on the blue side and a
300 lines mm$^{-1}$ grating on the red side.  A summary of the observations is shown in Table 1.

	The 1.5 m telescope observations were taken with a fixed N-S slit within a few hours
of culmination.  For all other systems, observations were taken with a long slit at the
parallactic angle.  Basic reduction steps were applied to the spectra using standard IRAF
routines.  Although every effort was made to minimize differential slit losses, in view of the
variable slit widths and seeing, we have not attempted to derive absolute spectrophotometry.  After
standard star calibration, we estimate that the relative spectrophotometric accuracy is $\sim$30\%,
based on comparisons of observations of individual targets at different epochs with different
instruments.  Spectra were corrected for telluric absorption, and all observations for
a given target were combined, weighted by S/N, to produce a final spectrum.  Sample spectra
are shown in Figure 4.

\begin{deluxetable}{ccccccc}
\tabletypesize{\scriptsize}
\tablecaption{Summary of CGRaBS observations.}
\tablewidth{0pt}
\tablehead{

  &
  &
  \colhead{Wavelength}&
  \colhead{Spectral}&
  \colhead{Typical}&
  \colhead{Typical}\\

  \colhead{Telescope}&
  \colhead{Dates}&
  \colhead{range}&
  \colhead{resolution}&
  \colhead{seeing}&
  \colhead{exposure}\\

  \colhead{}&
  \colhead{}&
  \colhead{(\AA)}&
  \colhead{(\AA)}&
  \colhead{(arcsec)}&
  \colhead{(s)}}

\startdata
9.2 m Hobby*Eberly&Ongoing, 2002--present&4100--9700&17&1.5&600, 1200\\[3pt]
\hline\\[-6pt]
&2005 May 27--31&4250--8250&12&1.5&600, 1200\\
2.7 m Harlan J. Smith&2005 Oct 27--31&4250--8250&12&2.0&600, 1200, 1800\\
&2007 Mar 26--28&4250--8250&12&3.0&600, 1200, 1800\\[3pt]
\hline\\[-6pt]
1.5 m CTIO&2005B&3500--9000&17&1.5&1200, 1800\\[3pt]
\hline\\[-6pt]
\multirow{2}{*}{3.6 m NTT}&2006 Aug 29--Sep 1&3900--9100&10&1.3&600, 1800\\
&2007 Jan 22--25&3900--9100&10&1.0&600, 1200\\[3pt]
\hline\\[-6pt]
\multirow{4}{*}{5 m Hale}&2005 Nov 5--9&3300--9500&5\tablenotemark{a}, 16\tablenotemark{b}&2.5&600, 1200\\
&2006 Aug 17--18&3300--9500&5\tablenotemark{a}, 16\tablenotemark{b}&1.7&600, 1200\\
&2007 Jan 15--16&3300--9500&5\tablenotemark{a}, 16\tablenotemark{b}&2.0&600, 1200\\
&2007 Apr 19--21&3300--9500&5\tablenotemark{a}, 16\tablenotemark{b}&2.5&600, 1200\\[3pt]
\hline\\[-6pt]
8.2 m VLT-Kueyen&Period 78&3500--8000&17&1.2&600, 900, 1200, 1800\\[3pt]
\hline\\[-6pt]
&2006 Jul 22--23&3300--9300&3\tablenotemark{a}, 11\tablenotemark{b}&1.5&600, 1200\\
10 m Keck I&2006 Oct 28&3300-9300&3\tablenotemark{a}, 11\tablenotemark{b}&1.5&600, 1200\\
&2006 Nov 24&3600-9600&3\tablenotemark{a}, 11\tablenotemark{b}&2.5&600, 1200\\
\enddata

\tablenotetext{a}{Blue side.}
\tablenotetext{b}{Red side.}

\end{deluxetable}

\begin{figure}[h]
\centering
\epsscale{0.795}
\plotone{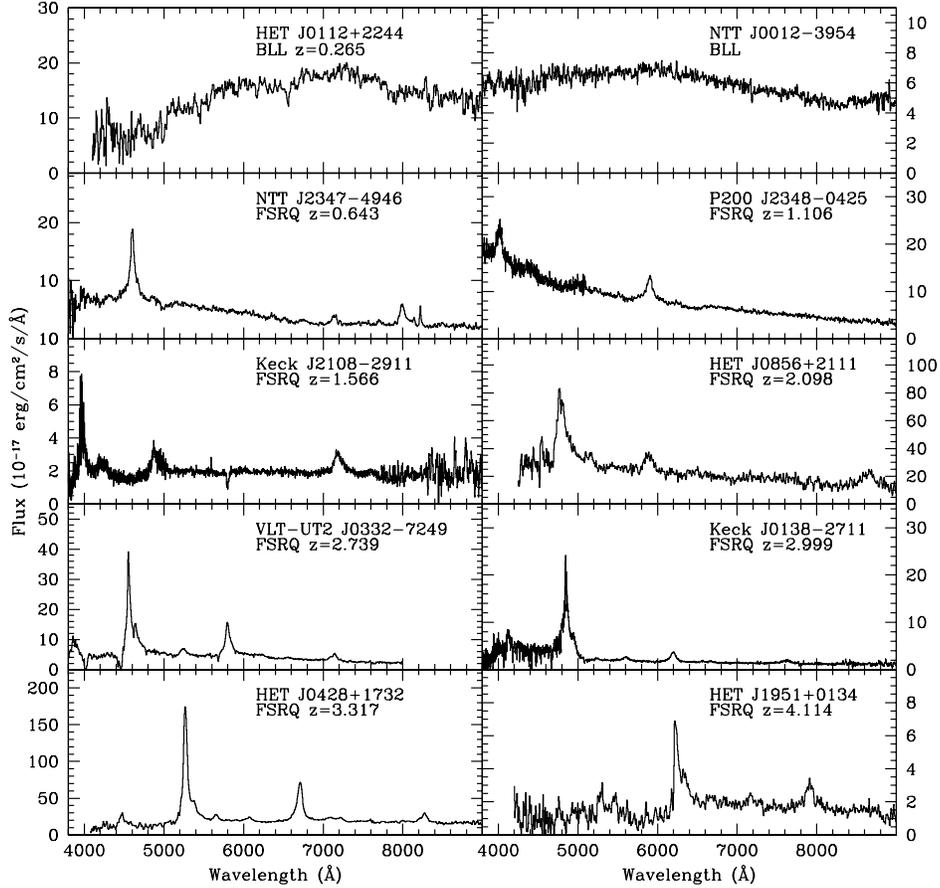}
\label{sampspec}
\caption[sampspec]{Sample CGRaBS spectra.}
\end{figure}

\subsubsection{Results}

	Our spectral analysis starts with a basic source classification.  The vast majority
($84$\%) are flat-spectrum radio quasars (FSRQs) dominated by strong broad emission lines. The weak-lined
BL Lac (BLL) class is somewhat heuristically defined; here we designate as BLLs sources
that exhibit the following properties \citep{marcha}: (1) emission line equivalent width $<5$
\AA, and (2) H/K 4000 \AA\ break contrast $\equiv (f^+ - f^-)/f^+ < 0.4$, where $f^+$ ($f^-$) is the
flux density redward (blueward) of the break.  It is often possible to establish that a source is a BLL
even when the redshift is impossible to determine.  For sources with $R>15$, we compute $M_R$ for the
$\Lambda$CDM concordance cosmology (smaller $R$ values are usually host-dominated, in any case) and
classify broad emission line sources with $M_R>-23$ as AGN.  Thus, we list here three blazar
designations: continuum-dominated BLLs, low-luminosity broad-line AGN, and luminous broad-line FSRQs.
A small number of non-blazar sources is also present.  Sources with narrow lines ($v_\mathrm{FWHM} <
1000$ km s$^{-1}$) are denoted as narrow-line radio galaxies (NLRGs).  Sources with small line
equivalent widths but large H/K break contrasts are denoted as galaxies. These low-redshift sources may
represent the low-luminosity extension of the blazar phenomenon.  One extremely compact planetary nebula
(PN) made our survey cuts.  Finally, in four cases, the radio position was within
$1.5\arcsec$ of a field star that dominated the initial spectrum.  With improved imaging,
the fainter CGRaBS blazar counterparts can be identified.

	Redshifts were measured by cross-correlation analysis.  For a modest number of FSRQs,
only a single strong, broad emission line is identified.  In most cases, we conservatively identify this as
Mg {\sc ii} $\lambda 2800$, supported by the absence of strong lines expected for other
identifications and, often, by Fe {\sc ii} structure in the surrounding continuum.  Nevertheless, these
redshifts are flagged by a colon (:), indicating possible systematic uncertainty.  Absorption line
redshifts were obtained for some BLLs.  In a few
cases, the BLL sources had multiple observations, and we were able to obtain emission line
redshifts when the source was in a {\it low} continuum state and the emission line
equivalent widths were relatively large.  A few additional BLLs have redshift constraints,
with upper limits from the lack of Ly-$\alpha$ absorption in the UV and lower limits from
clearly identified (typically Mg {\sc ii}) intergalactic absorption systems. We have also measured
continuum flux densities and equivalent and kinematic widths for the strong optical/UV resonance
lines.  These will be used to study the black hole masses and evolution.

	Table 2 presents the first page of the CGRaBS catalog; the full table appears in the online
edition.  Here we include the precise position, the 8.4 GHz core flux density, the FoM,
the $R$ magnitude, the extinction $A_R$, and the optical classification and redshift, if any.

\begin{deluxetable}{crrrrccccc}
\tabletypesize{\scriptsize}
\tablewidth{0pt}
\tablecaption{The CGRaBS catalog.}
\tablehead{

 &&&\colhead{$S_{8.4\,\mathrm{GHz}}$}&&\colhead{X-ray}&\colhead{$R$}&\colhead{$A_R$}&\\

 \colhead{Name}&
 \colhead{R.A.\tablenotemark{a}}&
 \colhead{Decl.\tablenotemark{a}}&
 \colhead{(mJy)}&
 \colhead{FoM}&
 \colhead{flag\tablenotemark{b}}&
 \colhead{(mag)}&
 \colhead{(mag)}&
 \colhead{Type\tablenotemark{c}}&
 \colhead{$z$}}

\startdata

J0001$-$1551&00 01 05.33&$-$15 51 07.1&335.90&0.050&$-$&18.09&0.08&FSRQ&2.044\\
J0001$+$1914&00 01 08.62&$+$19 14 33.8&504.20&0.105&$-$&20.50&0.11&FSRQ&3.100\\
J0003$+$2129&00 03 19.35&$+$21 29 44.4&269.70&0.098&$-$&19.75&0.12&AGN&0.450\\
J0004$-$1148&00 04 04.92&$-$11 48 58.4&774.90&0.112&$-$&19.09&0.08&BLL&\\
J0004$+$4615&00 04 16.13&$+$46 15 18.0&214.80&0.060&$-$&20.44&0.24&FSRQ&1.810\\
J0004$+$2019&00 04 35.76&$+$20 19 42.2&162.50&0.058&$-$&20.25&0.10&BLL&0.677\\
J0004$-$4736&00 04 35.68&$-$47 36 18.6&780.40&0.067&$-$&15.88&0.05&FSRQ&0.880\\
J0005$-$1648&00 05 17.93&$-$16 48 04.7&281.70&0.050&$-$&18.37&0.07&&\\
J0005$+$0524&00 05 20.21&$+$05 24 10.7&228.90&0.066&$-$&16.26&0.08&FSRQ&1.900\\
J0005$+$3820&00 05 57.18&$+$38 20 15.2&1077.60&0.137&$-$&17.16&0.24&FSRQ&0.229\\
J0006$-$0623&00 06 13.89&$-$06 23 35.3&3296.90&0.135&$-$&17.14&0.10&BLL&0.347\\
J0006$+$2422&00 06 48.79&$+$24 22 36.5&230.90&0.049&$-$&18.71&0.24&FSRQ&1.684\\
J0008$-$2339&00 08 00.37&$-$23 39 18.1&377.20&0.057&$-$&16.16&0.05&FSRQ&1.410\\
J0008$-$4619&00 08 37.54&$-$46 19 40.8&176.00&0.041&$-$&16.51&0.04&FSRQ&1.850\\
J0010$+$2047&00 10 28.74&$+$20 47 49.7&272.10&0.071&$-$&18.41&0.23&FSRQ&0.600\\
J0010$+$1058&00 10 31.01&$+$10 58 29.5&245.00&0.141&$-$&12.22&0.26&AGN&0.089\\
J0010$+$1724&00 10 33.99&$+$17 24 18.8&867.50&0.069&$-$&16.90&0.10&FSRQ&1.601\\
J0010$-$3027&00 10 35.75&$-$30 27 47.4&316.70&0.050&$-$&19.07&0.04&FSRQ&1.189\\
J0010$-$2157&00 10 53.65&$-$21 57 04.2&358.90&0.049&$-$&19.68&0.06&&\\
J0011$-$2612&00 11 01.25&$-$26 12 33.4&520.40&0.125&$-$&19.64&0.05&FSRQ&1.093\\
J0011$+$0057&00 11 30.40&$+$00 57 51.8&278.70&0.072&$-$&20.06&0.07&FSRQ&1.492\\
J0012$+$3353&00 12 47.38&$+$33 53 38.5&213.40&0.075&$-$&20.40&0.14&FSRQ&1.682\\
J0012$-$3954&00 12 59.91&$-$39 54 25.8&1554.20&0.181&$-$&18.05&0.03&BLL&\\
J0013$-$1513&00 13 20.71&$-$15 13 47.9&202.40&0.055&$-$&19.15&0.06&FSRQ&1.838\\
J0013$-$0423&00 13 54.13&$-$04 23 52.3&345.50&0.059&$-$&19.89&0.08&FSRQ&1.075\\
J0013$+$1910&00 13 56.38&$+$19 10 41.9&393.70&0.110&$-$&18.17&0.13&BLL&\\
J0015$-$1812&00 15 02.49&$-$18 12 50.9&378.30&0.054&$-$&19.65&0.09&FSRQ&0.743\\
J0016$-$0015&00 16 11.09&$-$00 15 12.5&732.50&0.040&$-$&19.72&0.08&FSRQ&1.574\\
J0017$+$8135&00 17 08.48&$+$81 35 08.1&1361.10&0.140&$-$&15.95&0.49&FSRQ&3.387\\
J0017$-$0512&00 17 35.82&$-$05 12 41.7&225.20&0.050&$-$&17.60&0.08&FSRQ&0.227\\
J0019$+$2021&00 19 37.85&$+$20 21 45.6&1232.90&0.098&$-$&19.70&0.16&BLL&\\
J0019$-$3031&00 19 42.67&$-$30 31 18.6&485.40&0.058&$-$&19.64&0.06&FSRQ&2.677\\
J0019$+$2602&00 19 39.78&$+$26 02 52.3&458.50&0.046&X&15.04&0.08&FSRQ&0.284\\
J0019$+$7327&00 19 45.79&$+$73 27 30.0&1330.70&0.094&$-$&18.26&0.86&FSRQ&1.781\\
J0022$+$4525&00 22 06.61&$+$45 25 33.8&307.50&0.043&$-$&20.72&0.19&FSRQ&1.897\\
J0022$+$0608&00 22 32.44&$+$06 08 04.2&301.20&0.052&$-$&19.07&0.06&BLL&\\
J0023$+$4456&00 23 35.44&$+$44 56 35.8&240.00&0.065&$-$&21.70&0.16&FSRQ&1.062\\
J0024$+$2439&00 24 27.33&$+$24 39 26.3&188.00&0.040&$-$&19.20&0.08&FSRQ&1.444\\
J0025$-$2227&00 25 24.25&$-$22 27 47.6&248.80&0.052&$-$&18.73&0.04&&\\
J0026$-$3512&00 26 16.39&$-$35 12 48.7&314.70&0.108&$-$&19.68&0.03&&\\
J0027$+$2241&00 27 15.37&$+$22 41 58.2&323.80&0.055&$-$&15.60&0.10&FSRQ&1.108\\

\enddata

\tablenotetext{a}{J2000.0 position.}
\tablenotetext{b}{``X'' indicates that a source would not satisfy the FoM cutoff if its X-ray flux were ignored.  See \S2.}
\tablenotetext{c}{See \S\S3.2.1-2 for discussion of the type classifications.}

\end{deluxetable}

\section{Discussion}

To date, we have 1226 redshifts and 64 BLLs with unknown redshift.  Thus classification is
79\% complete with respect to the entire survey and 85\% for objects with known $R < 23$.  
So far, 10.3\% of all
objects classified are BLLs, 3.4\% are AGN, and 1\% are NLRGs.  Figure 5 shows the completeness as a 
function of magnitude.  Source classification and redshifts are $>$85\% complete to $R=20$.
While the completeness
drops off rapidly beyond this, so do the source counts, and so reaching $>$95\% completeness
at the survey limit is feasible.  Note, however, that only $\sim$52\% of the
BLLs have redshifts and that this fraction falls off quickly above $R=18$. Clearly, pushing the
largely complete BLL sample fainter than $R=20$ will be a challenge.

\begin{figure}[ht]
\centering
\includegraphics[width=0.7\textwidth,trim=0in 0.7in 0in 0in]{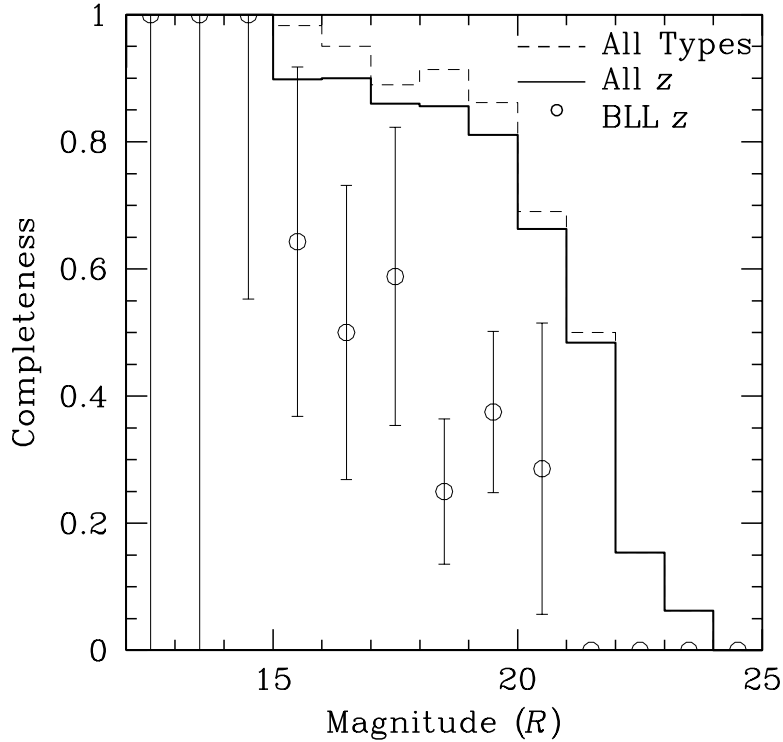}
\label{compltype}
\caption[compltype]{Completeness as a function of $R$ for source identification and redshift
measurement (histograms). The points show the fraction of identified BLL with measured
redshift; small numbers lead to substantial error bars.
}
\end{figure}

	We defer full discussion of the sample properties until we reach our expected
95\% completeness to $R=23$.  However, it is already interesting to examine the redshift
distribution of the sources detected to date (Figure 6).  The non-BLL (largely
FSRQ) distribution peaks at
$z\approx 1.3$ and has an exponential fall-off (${\rm d}N/{\rm d}z \propto 10^{-0.6z}$)
to high redshift, extending to $z=5.5$.
From SED information on optically faint sources, we expect the high-redshift population
to increase somewhat in the complete CGRaBS sample, but it is clear that there will be
only a handful of radio-bright blazars at $z>4$. If any of these are detected by the LAT,
as expected, they will be particularly important targets for multiwavelength spectral
and variability studies.  In fact, with only $\sim$40 sources at $z>3$, careful study
of these few high-redshift objects will be important for several LAT programs, e.g.,
extragalactic background light (EBL) studies and studies of jet evolution and interaction
with the CMBR.

\begin{figure}[t]
\centering
\includegraphics[width=0.7\textwidth,trim=0in 0.7in 0in 0in]{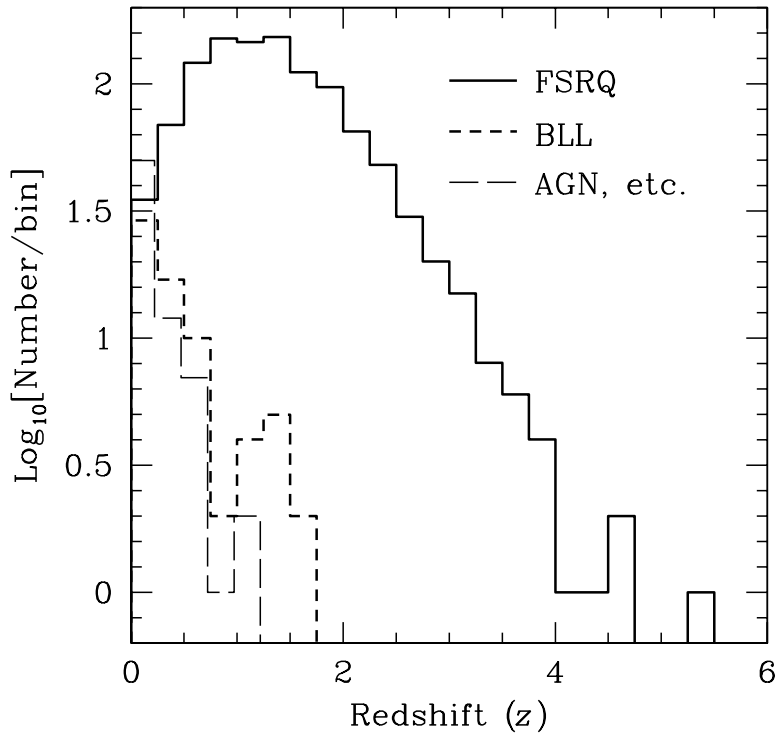}
\label{zdist}
\caption[zdist]{Redshift distributions for the (partly complete) CGRaBS survey. The solid-line histogram
shows FSRQs. The short-dashed histogram gives the redshift distribution for solved BLLs.
The long-dashed histogram shows a variety of other AGN (NLRGs, passive ellipticals, etc.),
which contribute only at very low redshift.}
\end{figure}

	We are also assembling an important new sample of radio-bright BLLs. To date,
we have 133 sources definitively classified as BLLs, but this will likely grow since
a substantial number of other sources have observed BLL-like spectra but need somewhat
improved S/N observations to exclude emission lines with EW $\ge$ 5 \AA\ throughout the observed
spectrum. Among the brighter sources $\sim$15\% are BLL; at this incidence, we expect 
$\sim$245 sources to have a final BLL
classification. As noted, it will be very tough to obtain redshifts of the faintest
BLLs. However, the 70 redshifts already in hand represent a substantial radio-bright
sample. For example, it is twice the size of the 1 Jy sample \citep{stet91}
and extends to nearly twice the redshift. At present, we have 11 BLLs at $z>1$, a 
third of all known $z>1$ BLLs, so that the full survey should be useful for probing
evolution of this population.

	Of course, the most important application of the CGRaBS catalog is the identification
with other all-sky samples and the generation of multiwavelength SEDs. We are already
examining the radio to X-ray spectra of these sources and eagerly look forward to
the upcoming sky surveys with {\it AGILE}, the air-\v{C}erenkov TeV observatories, and
especially {\it GLAST}, which will measure the $\gamma$-ray power peak expected for many of these
sources.

\acknowledgments

S.\ E.\ H.\ was supported by SLAC under DOE contract DE-AC03-76SF00515.

The Hobby*Eberly Telescope (HET) is a joint project of the University of 
Texas at Austin, the Pennsylvania State University, Stanford University, 
Ludwig-Maximilians-Universit\"{a}t M\"{u}nchen, and 
Georg-August-Universit\"{a}t G\"{o}ttingen.  The HET is named in honor 
of its principal benefactors, William P.\ Hobby and Robert E.\ Eberly.

The Marcario Low Resolution Spectrograph is named for Mike Marcario of 
High Lonesome Optics, who fabricated several optics for the instrument 
but died before its completion.  The LRS is a joint project of the 
Hobby*Eberly Telescope partnership and the Instituto de Astronom\'{i}a 
de la Universidad Nacional Aut\'{o}noma de M\'{e}xico.

This paper includes data taken at the McDonald Observatory of the 
University of Texas at Austin.

The data in this paper are based partly on observations obtained at the 
Hale Telescope, Palomar Observatory, as part of a collaborative 
agreement between the California Institute of Technology, its divisions 
Caltech Optical Observatories and the Jet Propulsion Laboratory 
(operated for NASA), and Cornell University.

Some of the data presented herein were obtained at the W.\ M.\ Keck 
Observatory, which is operated as a scientific partnership among the 
California Institute of Technology, the University of California and the 
National Aeronautics and Space Administration. The Observatory was made 
possible by the generous financial support of the W.\ M.\ Keck 
Foundation.

The authors wish to recognize and acknowledge the very significant
cultural role and reverence that the summit of Mauna Kea has always
had within the indigenous Hawaiian community.  We are most fortunate
to have the opportunity to conduct observations from this mountain.

Some of the data in this paper are based on observations made with ESO
telescopes at the La Silla Observatory under program 077.B-0056(A) and
ID 078.B-0275(B).

Funding for the SDSS and SDSS-II has been provided by the Alfred P.\ 
Sloan Foundation, the Participating Institutions, the National Science 
Foundation, the U.\ S.\ Department of Energy, the National Aeronautics 
and Space Administration, the Japanese Monbukagakusho, the Max Planck 
Society, and the Higher Education Funding Council for England.  The SDSS 
Web site is {\tt http://www.sdss.org/}.

The SDSS is managed by the Astrophysical Research Consortium for the 
Participating Institutions.  The Participating Institutions are the 
American Museum of Natural History, Astrophysical Institute Potsdam, 
University of Basel, University of Cambridge, Case Western Reserve 
University, University of Chicago, Drexel University, Fermilab, the 
Institute for Advanced Study, the Japan Participation Group, Johns 
Hopkins University, the Joint Institute for Nuclear Astrophysics, the 
Kavli Institute for Particle Astrophysics and Cosmology, the Korean 
Scientist Group, the Chinese Academy of Sciences (LAMOST), Los Alamos 
National Laboratory, the Max Planck Institute for Astronomy (MPIA), the 
Max Planck Institute for Astrophysics (MPA), New Mexico State 
University, Ohio State University, University of Pittsburgh, University 
of Portsmouth, Princeton University, the United States Naval 
Observatory, and the University of Washington.

This research has made use of the NASA/IPAC Extragalactic Database (NED), which
is operated by the Jet Propulsion Laboratory, California Institute of Technology,
under contract with the National Aeronautics and Space Administration.


\begin{thebibliography}{}
\bibitem[Adelman-McCarthy et al.(2007)]{sdss}Adelman-McCarthy, J.\ et al.\ 2007, ApJS, in press.
\bibitem[Bock et al.(1999)]{sumss}Bock, D.\ C.-J., Large, M.\ I., \& Sadler, E.\ M.\ 1999, AJ, 117, 1578.
\bibitem[Condon et al.(1998)]{nvss}Condon, J.\ J.\ et al.\ 1998, AJ, 115, 1693.
\bibitem[Hartman et al.(1993)]{hart93}Hartman, R.\ C.\ et al.\ 1993, ApJ, 407, L41.
\bibitem[Hartman et al.(1999)]{3eg}Hartman, R.\ C.\ et al.\ 1999, ApJS, 123, 79.
\bibitem[Healey et al.(2007)]{crates}Healey, S.\ E.\ et al.\ 2007, ApJS, 171, 61.
\bibitem[Hill et al.(1998)]{lrs}Hill, G.\ J.\ et al.\ 1998, SPIE, 3355, 375.
\bibitem[Kniffen et al.(1993)]{kniffen}Kniffen, D.\ A.\ et al.\ 1993, ApJ, 411, 133.
\bibitem[March\~{a} et al.(1996)]{marcha}March\~{a}, M.\ J.\ M.\ et al.\ 1996, MNRAS, 281, 425.
\bibitem[Massaro et al.(2007)]{asdc}Massaro, E.\ et al.\ 2007, {\tt http://www.asdc.asi.it/bzcat/}
\bibitem[Mattox et al.(2001)]{mattox}Mattox, J.\ R.\ et al.\ 2001, ApJS, 135, 155.
\bibitem[Monet et al.(2003)]{usno}Monet, D.\ G.\ et al.\ 2003, AJ, 125, 984.
\bibitem[Myers et al.(2003)]{class}Myers, S.\ T.\ et al.\ 2003, MNRAS, 341, 1.
\bibitem[Padovani \& Giommi(1995)]{hbl}Padovani, P.\ \& Giommi, P.\ 1995, ApJ, 444, 567.
\bibitem[Schneider et al.(2007)]{sdssq}Schneider, D.\ P.\ et al.\ 2007, AJ, 134, 102.
\bibitem[Schlegel, Finkbeiner \& Davis(1998)]{schleg}Schlegel, D.\ J., Finkbeiner, D.\ P.\ \& Davis, M.\ 1998, ApJ, 500, 525.
\bibitem[Shetrone et al.(2007)]{shetrone}Shetrone, M. et al.\ 2007, PASP, 119, 556.
\bibitem[Sowards-Emmerd et al.(2003)]{srm03}Sowards-Emmerd, D.\ et al.\ 2003, ApJ, 590, 109.
\bibitem[Sowards-Emmerd et al.(2005)]{srm05}Sowards-Emmerd, D.\ et al.\ 2005, ApJ, 626, 95.
\bibitem[Stickel et al.(1991)]{stet91}Stickel, M.\ et al.\ 1991, ApJ, 374, 431.
\bibitem[Turriziani et al.(2007)]{roxa}Turriziani, S.\ et al.\ 2007, astro-ph/0705.1498.
\bibitem[V\'{e}ron-Cetty \& V\'{e}ron(2006)]{veron}V\'{e}ron-Cetty, M.-P.\ \& V\'{e}ron, P.\ 2006, A\&A, 455, 773.
\bibitem[Voges et al.(1999)]{rass}Voges, W.\ et al.\ 1999, A\&A, 349, 389.
\end{thebibliography}
\end{document}